\documentclass[iopams]{iopart}
\usepackage{geometry}
\usepackage{graphics}
\usepackage{graphicx}
\usepackage{amssymb}
\usepackage{color}

\begin{document}
\title[Asymptotic approximant for the bending angle of light]{An asymptotically consistent approximant for the equatorial bending angle of light due to Kerr black holes}
\author{Nathaniel S. Barlow$^1$, Steven J. Weinstein$^2$, Joshua A. Faber$^{1,3}$}
\address{$^1$ School of Mathematical Sciences, Rochester Institute of Technology, Rochester, NY 14623} 
\address{$^2$ Department of Chemical Engineering, Rochester Institute of Technology, Rochester, NY 14623} 
\address{$^3$ Center for Computational Relativity and Gravitation, Rochester Institute of Technology, Rochester, NY 14623} 
\ead{nsbsma@rit.edu}

\begin{abstract}
An accurate closed-form expression is provided to predict the bending angle of light as a function of impact parameter for equatorial orbits around Kerr black holes of arbitrary spin. This expression is constructed by assuring that the weak- and strong-deflection limits are explicitly satisfied while maintaining accuracy at intermediate values of impact parameter via the method of asymptotic approximants (Barlow et al, 2017 \textit{Q.~J.~Mech.~Appl.~Math.}, 70 (1): 21-48).  To this end, the strong deflection limit for a prograde orbit around an extremal  black hole is examined, and the full non-vanishing asymptotic behavior is determined.  The derived approximant may be an attractive alternative to computationally expensive elliptical integrals used in black hole simulations.  \\

\end{abstract}

\submitto{\CQG}
\maketitle

\section{Introduction}\label{sec:intro}
Among the many and varied ways in which Einstein's general theory of relativity modified our understanding of gravity, the deflection of light by gravitating objects was perhaps the earliest-tested, having been measured during the solar eclipse of 1919 by Eddington, with gravitational lensing now its own recognized field of astronomy (see \cite{Bozza:2009yw} for a review of strong-field lensing by black holes).  The properties of light propagation have been, up to the direct detection of gravitational waves in the past year \cite{Abbott:2016blz,Abbott:2016nmj}, the source of our most direct information of the geometry of dynamical spacetimes, and underlie some of the most important tests to date on general relativity and its predicted consequences.   For example, the various relativistic time delays introduced into the signal from pulsars in relativistic binaries \cite{Goicoechea:1992zz} like PSR B1913+16 \cite{Weisberg:2016jye} and PSR J0737-3039 \cite{Kramer:2006nb}, or those that would be induced on the signals for pulsars orbiting supermassive black holes \cite{Wang:2009yp}, could  provide extremely tight constraints on certain  post-Newtonian model parameters that characterize many modified theories of gravity (see \cite{Will:2014kxa,Berti:2015itd} for  thorough reviews).
  
 Given the centrality of light propagation to the field of astrophysics, a significant amount of work has gone into classifying the null geodesics that describe photon trajectories in the presence of either a Schwarzschild (non-spinning) or Kerr (spinning) black hole (hereafter BH).  For the case of generic orbits around the former, and equatorial ones around the latter, it has long been known that there exists an innermost circular orbit (ICO; not to be confused with the innermost stable circular orbit [ISCO] that applies to massive objects traveling on timelike geodesics), which represents a case of infinite deflection angle for a photon.  The strong deflection limit, for which photons travel close to the ICO, was first explored in 1930 by Hagihara \cite{Hagihara:1930zz}, and further elucidated by Darwin \cite{Darwin:1959zz} in the 1950's, while results for Kerr geodesics were derived by Carter in the 1960's \cite{Carter:1968rr}.  For a thorough description of all of these results, the review in \cite{Chandrasekhar:1983book} remains an essential reference.
 The properties of the deflection angle in both the weak- and strong-field regimes were studied for Schwarzschild \cite{Iyer:2006cn} and Kerr \cite{Iyer:2009wa,Iyer:2009hq} black holes by Iyer and collaborators, and later extended by others to the near-equatorial case \cite{Chakraborty:2015aua}.

For Kerr BHs of all spins (including for the Schwarzschild case), the exact closed-form solutions for the bending angle of light are expressed in terms of elliptic integrals, effectively rendering them a computational bottleneck in applications that require large numbers of trajectories to be calculated in a short amount of time.  In particular, this situation occurs during numerical ray-tracing of the emission from objects like accretion disks, or detailed calculations of lensing patterns in the strong-field regime.  The asymptotic properties of the weak-field limit are well-understood, as are those in the extreme strong-field regime \cite{Bozza:2005tg,Bozza:2006nm}, and approximations exist to handle various cases for photons passing through the latter \cite{Beloborodov:2002mr}.   Further complications are introduced in extensions of the Schwarzschild or Kerr metrics, such as those that introduce a cosmological constant term (Kerr-de Sitter metric) \cite{Kraniotis:2005zm,Kraniotis:2010gx,Bhadra:2010jr,Chakraborty:2016hsk,Charbulak:2017bpj}, or both a cosmological constant and an electric charge (Kerr-Newman-de Sitter metric) \cite{Stuchlik:2008xk,Kraniotis:2014paa}.  Complete geodesic solutions are known in terms of Jacobi integrals (see \cite{Cadez:2004cg,Munoz:2014zz} for expressions involving Schwarzschild spacetimes, and \cite{Cadez:1998zz} for those involving Kerr), which are also a significant computational bottleneck in simulations, and are not easily invertible.  Approximate schemes to construct geodesics remain a viable area of research, since accurate and efficient approximation schemes can be included in ray-tracing codes (see, e.g.,  \cite{Semerak:2014kra} for an approximate scheme describing Schwarzschild geodesics) and numerical work concerned primarily with gravitational lensing often make extensive use of approximate schemes \cite{Aazami:2011tu,Aazami:2011tw}.  The difficulties in using exact closed-form expressions are evident in ray-tracing codes like GYOTO \cite{Vincent:2011wz} and GRay \cite{Chan:2013zz} that rely primarily on numerical integration techniques.  The GeoKerr-YNOGK  code \cite{Dexter:2009fg,Yang:2013yoa} relies heavily on elliptic integrals that also must be evaluated numerically.

In the relatively near future, astronomers are likely to make dramatic gains in our ability to directly image the strong-field regions surrounding black holes, particularly through ground-based interferometry projects like the Event Horizon Telescope \cite{Doeleman:2008qh} .  While high-angle light bending in the strong-field regime may be difficult to observe directly \cite{Virbhadra:1999nm}, understanding how bending affects the global emission we see from black hole accretion (and what we may infer about the black hole's spin and the spacetime geometry) will become an even more important concern.  Indeed, future telescopes that are sensitive to physical effects, rather than purely those due to geometrical optics, may be used to study the spins of BHs \cite{Tamburini:2011tk}.

Here, we consider the simplest aspect of light deflection, that of equatorial orbits around Kerr black holes for photon trajectories that begin and end at large separations from the black hole.  We first clarify and extend the work of \cite{Iyer:2006cn,Iyer:2009wa,Iyer:2009hq}, to provide mathematically well-posed expressions for the asymptotic behavior of light bending in the strong-field regime.  To do so we also account for the special cases of Schwarzschild (zero spin) and extremal Kerr (extremal spin) BHs.  Unlike previous schemes that use numerical fitting, we are able to construct bending angle formulae that a.) describe  the strong-field, weak-field, and all intermediate cases to high accuracy; and b.) may be generated to arbitrary orders of accuracy without any recourse to ad hoc assumptions or numerical curve-fitting.
The methodology used here falls under the class of \textit{asymptotic approximants}, a new method that has seen recent success in the description of thermodynamic phase behavior~\cite{Barlow:2012, Barlow:2014, Barlow:2015} and the solution of nonlinear boundary value problems~\cite{Barlow:2017}.  Asymptotic approximants are closed-form expressions that behave asymptotically like the exact solution of a problem towards one region while having a Taylor expansion (about a different region) that matches the expansion of the exact solution of the problem to a specified order.  The well-known Pad\'e approximant, which is the quotient of two polynomials, may be thought of as a subset of asymptotic approximants.  Here, we instead choose asymptotic approximants that contain the appropriate blend of logarithmic and power-law terms to mimic the asymptotic structure of the bending angle of light.

Our paper is organized as follows: In Section~\ref{sec:notation}, we lay out the basic equations governing null geodesics in Schwarzschild and Kerr spacetime, establish notation, and discuss the current understanding of light bending angles in both the strong-field and weak-field regimes as a function of the BH spin.  In Section~\ref{sec:asymptotics}, we describe the asymptotic approximant, provide formulae for obtaining its coefficients, and show the results at various orders of the approximant for Kerr BHs of all spins, including Schwarzschild and extremal Kerr BHs.  Finally, in Section~\ref{sec:conclusions}, we discuss how these results may be extended to describe further dimensions of phase space as well as additional quantities relevant to photon propagation.  ~\ref{sec:a=1} provides the asymptotic analysis of a prograde orbit around an extremal Kerr black hole in its strong deflection limit.

\section{Light bending in the weak- and strong-field limits}\label{sec:notation}
\begin{figure*}[h!]
\begin{center}
\includegraphics[width=5in]{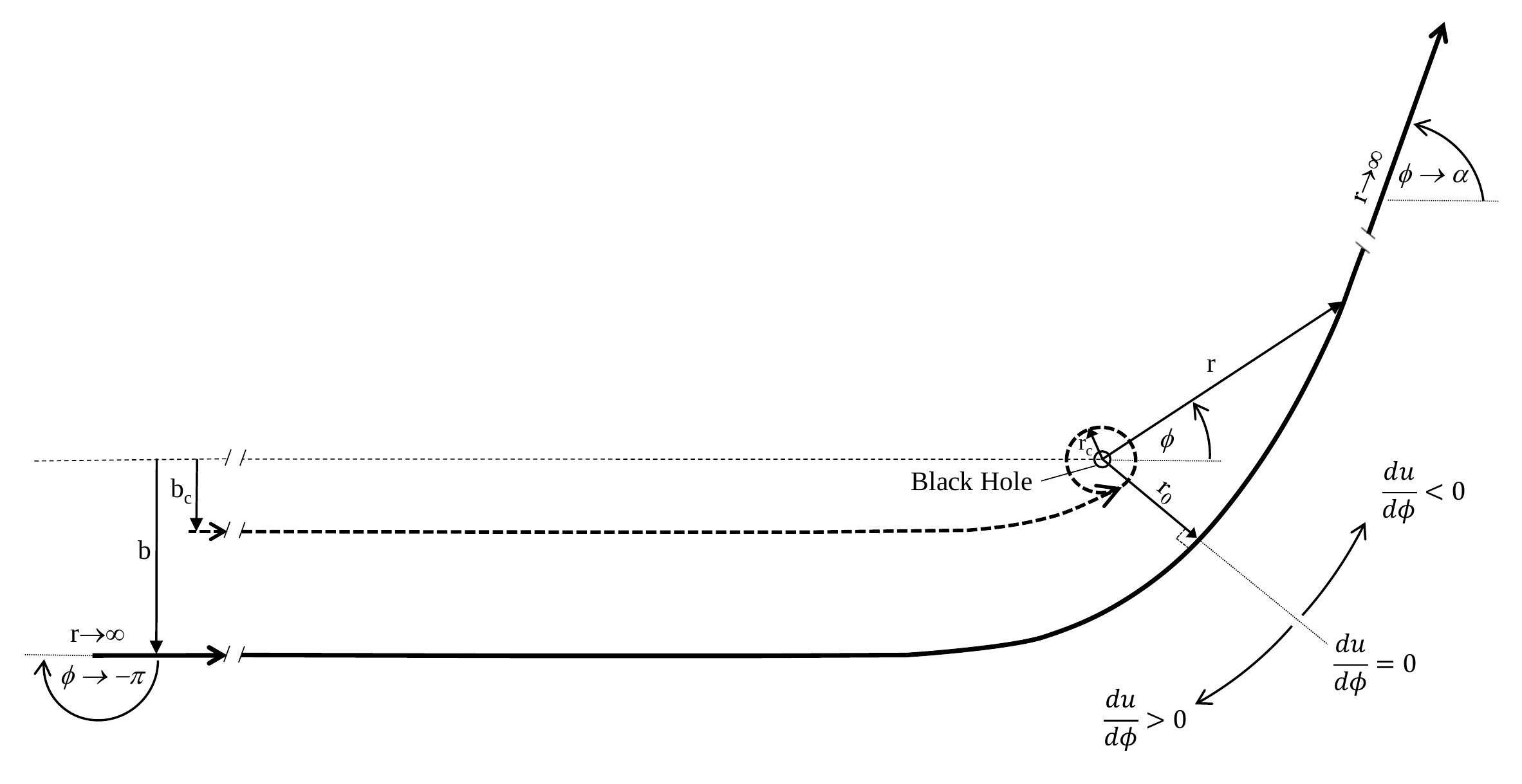}
\end{center}
\caption{Schematic of equatorial light bending, showing the various distances, angles, and overall orientation.  The solid line depicts an arbitrary trajectory of photon deflection, and shows the relationship between the impact parameter, $b$, the minimum radius, $r_0$, and the bending angle, $\alpha$.  Also shown is the Innermost Circular Orbit where $\alpha$ becomes infinite, showing the relationship between the critical impact and radius, $b_c$ and $r_c$, respectively.  The derivative $du/d\phi$, where $u=1/r$, is used in the interpretation of~(\ref{eq:dudphi}).}
\label{fig:schematic}
\end{figure*}

To introduce notation and explain the steps that lead to our new results, we briefly review the established theory for calculating light deflection in the equatorial plane of Kerr BH spacetimes, a thorough discussion of which can be found in \cite{Chandrasekhar:1983book} and numerous other references.  We begin from the Kerr spacetime metric \cite{Kerr:1963ud}, expressed in Boyer-Lindquist coordinates, centered at the black hole, letting the polar angle $\theta=\pi/2$ everywhere and suppressing all terms involving $d\theta$.   Under these assumptions, the line element $ds$ takes the form
\begin{eqnarray*}
ds^2 = -\left(1-\frac{2R}{r}\right)dt^2+\left(1-\frac{2R}{r}\right)^{-1}dr^2+\left(r^2+S^2+\frac{2RS^2}{r}\right)d\phi^2-\frac{4RS}{r}dt~d\phi
\end{eqnarray*}
in these coordinates, where  $t$ and $r$ are the coordinate time and radius, respectively, and $R\equiv GM_{\rm BH}/c^2$ is the gravitational radius of the BH  (of mass $M_{\rm BH}$), which may be set to unity if we choose it to be our distance unit (we follow standard relativistic convention where the gravitational constant $G$ and speed of light $c$ are also set to unity).  The parameter $\phi$ is the azimuthal angle, as depicted in Figure~\ref{fig:schematic}.  The spin parameter $S$ (with units of distance)  is defined as $S\equiv J_{\rm BH}/M_{\rm BH}c$, and the dimensionless spin parameter is $a\equiv S/R=J_{\rm BH}c/(GM_{\rm BH}^2)$, where $J_{\rm BH}$ is the BH's angular momentum.  An ``extremal'' black hole is one for which $|a|= 1$, or equivalently $|S|=R$.  The Kerr metric reduces to the familiar Schwarzschild form when $S=a=0$. 

To measure light deflection for photons in the equatorial plane passing by the black hole, we begin from the equations describing null geodesics, which may be parameterized (see, e.g., \cite{Chandrasekhar:1983book,Iyer:2009wa}) in terms of $u\equiv \frac{1}{r}$ and $\phi$ such that
\begin{eqnarray}\label{eq:dudphi}
\left|\frac{du}{d\phi}\right| = \frac{1-2u+a^2u^2}{1-2u\left(1-\frac{a}{b}\right)}\sqrt{2\left(1-\frac{a}{b}\right)^2 u^3 - \left(1-\frac{a^2}{b^2}\right)u^2+\frac{1}{b^2}}
\end{eqnarray}
where the impact parameter $b\equiv L/E$, and $L$ and $E$ are the $z$-component of angular momentum of the photon and its energy, respectively.  At large separations, the impact parameter plays its usual role, representing the coordinate distance between the BH center and the photon's closest approach in the absence of curvature due to the BH's gravity.  To resolve any ambiguities in the sign convention, we make the typical assumption that $b>0$, allowing the spin parameter to cover the full range $-1\le a\le 1$, where $a<0$ represents retrograde orbits and $a>0$ prograde ones.  

To calculate the bending angle $\alpha$ of a light ray passing by the black hole, we invert ~(\ref{eq:dudphi}), and follow the evolution from $u=0$ (i.e., $r\rightarrow\infty$) to the maximum value $u_0\equiv 1/r_0$, where $r_0$ is the distance of closest approach (see Figure~\ref{fig:schematic}), and then back to $u=0$, which doubles the value of the integral. Note that the $\pi$ term below arises from the unperturbed trajectory, as shown in Figure~\ref{fig:schematic} as $r\to\infty$ for $\frac{du}{d\phi}>0$.
\begin{eqnarray}\label{eq:alpha}
\alpha = -\pi + 2\int_0^{u_0} \frac{1-2u\left(1-\frac{a}{b}\right)}{[1-2u+a^2u^2]\sqrt{2\left(1-\frac{a}{b}\right)^2 u^3 - \left(1-\frac{a^2}{b^2}\right)u^2+\frac{1}{b^2}}} du.
\end{eqnarray}
When viewed as an initial value problem, one may take $\phi(u=0)=-\pi$ as an initial condition to ~(\ref{eq:dudphi}) and solve the problem assuming $d\phi/du>0$ (from $u=0$ to $u=u_0$) and $d\phi/du<0$ on the way back (from $u=u_0$ to $u=0$), as shown in Figure~\ref{fig:schematic}.
  
\subsection{The critical geodesic path}
The radius of closest approach to a BH, $r_0$, is dependent on the impact parameter, $b$, as shown in figure~\ref{fig:schematic}.  Note that there is a critical impact parameter $b_c$ and critical minimum distance $r_c$ coincident with the innermost circular orbit (ICO) radius (see figure~\ref{fig:schematic}), where the bending angle becomes infinite. These quantities are given by
\begin{eqnarray}
r_c(a) &=& 2+2\cos\left[\frac{2}{3}\cos^{-1}(-a)\right]\\
b_c (a)&=& 6\cos\left[\frac{1}{3}\cos^{-1}(-a)\right]-a
\end{eqnarray}
and are related through the expressions
\begin{eqnarray*}
r_c = 3\frac{b_c-a}{b_c+a} = \sqrt{\frac{b_c^2-a^2}{3}}.
\end{eqnarray*}
These parameters take the following values for Schwarzschild and extremal Kerr BHs (in units of $M_{\rm BH}$, as everywhere in this paper):
\begin{eqnarray*}
{\rm Schwarzschild}~ (a=0)&:&~~r_c = 3;~~b_c = 3\sqrt{3}\\
{\rm Extremal~Kerr, prograde}~ (a=1)&:&~~r_c = 1;~~b_c = 2\\
{\rm Extremal~Kerr, retrograde}~ (a=-1)&:&~~r_c = 4;~~b_c = 7
\end{eqnarray*}

More generally, there is a monotonic relationship between the impact parameter and distance of closest approach for a given BH spin, which may be found by setting the cubic term under the radical sign  in~(\ref{eq:dudphi}) to zero.  In terms of either $u_0$ or $r_0$, the relationship is given by\footnote{Note there is an error in the middle term of Eq.~16 and in the unnumbered equation prior to Eq.~20 in \cite{Iyer:2009wa}.}
\begin{eqnarray}\label{eq:ucubic}
0 = 2(b-a)^2 u_0^3 - (b^2-a^2)u_0^2+1;~~~~  0 = r_0^3-(b^2-a^2)r_0+2(b-a)^2.
\end{eqnarray}
The resulting  depressed cubic has the solution~\cite{Iyer:2009wa}
\begin{eqnarray}
r_0[b; a] = \frac{2}{\sqrt{3}}\sqrt{b^2-a^2}\cos\left\{\frac{1}{3}
\cos^{-1}\left(-3\sqrt{3}\frac{(b- a)^2}
{(b^2-a^2)^{3/2}}\right) \right\}.
\end{eqnarray}
We note that for prograde orbits around extremal Kerr BHs ($a=1$), the relationship between closest approach and impact parameter is significantly simpler, and reduces to $r_0 = b-1$.  This result is used to analyze the strong deflection limit for such a BH in~\ref{sec:a=1}.

We present the expressions that follow in terms of the perturbation parameter 
\begin{eqnarray}
\label{eq:impact}
b'\equiv1-\frac{b_c}{b}
\end{eqnarray}
such that $b'\to0$ corresponds to the strong deflection limit ($r_0\to r_c$) and $b'\to1$ to the weak deflection limit ($r_0\to \infty$).

\subsection{The weak deflection limit}
The expansion of~(\ref{eq:alpha}) about $b'=1$  is given by~\cite{Iyer:2009hq} (noting that $\frac{1}{b} = \frac{(1-b')}{b_c}$):
\begin{eqnarray}
\nonumber
\alpha=\sum_{n=1}^\infty a_n(b'-1)^n
\label{series}\\
\nonumber
a_1=-\frac{4}{b_c},~a_2=\frac{15\pi/4-4a}{b_c^2},~a_3=-\frac{128/3-10\pi a+4a^2}{b_c^3},\\~a_4=\frac{3465\pi/64-192a+285\pi/16a^2-4a^3}{b_c^4},~\dots
\label{eq:alpha1}
\end{eqnarray}

\subsection{The strong deflection limit}\label{sec:sdl}

The asymptotic behavior as $b'\to0$ (i.e. $b\to b_c$) of~(\ref{eq:alpha}) for $a\neq1$ is given by the expressions between Eqs.~20 and 21 in~\cite{Iyer:2009hq}, albeit using different notation; for $a=1$, this behavior is derived in~\ref{sec:a=1}.   For convenience, these results are combined (for all $a$), re-written in terms of $b'$, and  truncated consistently to a specific asymptotic order:
\begin{equation}
 \alpha\sim-\pi+\beta+\gamma\ln\zeta+\delta_{a,1}\frac{\sqrt{3}}{b'}-\gamma\ln b'+O(b'\ln b'),~~\delta_{a,1} = \left\{
     \begin{array}{ll}
    0 &:~ a\neq1\\
       1&:~ a=1
     \end{array}
   \right..
 \label{eq:alpha0}
\end{equation}
The constants in~(\ref{eq:alpha0}), whose functional dependence is determined entirely by the BH spin parameter $a$,  are defined as follows:
\begin{equation}
   \beta = \left\{
     \arraycolsep=1.4pt\def\arraystretch{2}
     \begin{array}{ll}
    0 &:~ a=0\\
       \frac{r_c^{5/2}[U_-V_-+U_+V_+]}{3\sqrt{(1-a^2)}[r_c^2-2r_c+a^2](1-a/b_c)} &:~ 0<|a|<1\\
       \frac{8\sqrt{3}-6}{9} &:~ a=-1\\
       \frac{\sqrt{3}-4}{3} &:~ a=1,
     \end{array}
   \right.
\label{eq:betaarray}
\end{equation}

\begin{equation}
   \gamma = \left\{
     \arraycolsep=1.4pt\def\arraystretch{2}
     \begin{array}{ll}
       \frac{2r_c^{3/2}\left[r_c-2\left(1-\frac{a}{b_c}\right)\right]}{\sqrt{3} [r_c^2-2r_c+a^2]\left(1-\frac{a}{b_c}\right)} &:~ -1\le a<1\\
       \frac{4}{3^{3/2}} &:~ a=1,
     \end{array}
   \right.
\label{eq:gammaarray}
\end{equation}

\begin{equation}
   \zeta = \left\{
  \arraycolsep=1.4pt\def\arraystretch{2}
     \begin{array}{ll}
       \frac{216(7-4\sqrt{3})}{\kappa} & :~-1\le a<1\\
       \frac{18}{2+\sqrt{3}} &:~ a=1.
     \end{array} 
   \right. 
\label{eq:zetaarray}
\end{equation}

\[U_\pm= \frac{3}{r_c}\left[\pm a^2 \mp 2\left(1-\frac{a}{b_c}\right)(1\pm\sqrt{1-a^2}) \pm r_c\left(1\pm\sqrt{1-a^2}-2\frac{a}{b_c}\right)\right],\]
\[V_\pm =\xi_\pm \ln\left[\frac{(1+\xi_{\pm})(1-\sqrt{3}\xi_\pm)}{(1-\xi_{\pm})(1+\sqrt{3}\xi_\pm)}\right],~ \kappa=b_c\frac{\left[3b_c\sqrt{b_c^2-a^2}-6\sqrt{3}(b_c-a)\right]}{(b_c^2-a^2)^{3/2}},\]
and
\[\xi_\pm= \sqrt{\frac{a^2}{a^2+2r_c(1\pm\sqrt{1-a^2})}}.\]
Note that, although piecewise notation is used above,~(\ref{eq:alpha0}) is a continuous function of $a$ for $a<1$.  The coefficient $\beta$ is only written as a piecewise function for $a<1$ so as to explicitly handle the removable singularities in the $0<|a|<1$ expression of $\beta$ at $a=0$ and $a=-1$.   For $a=1$, a different structure arises for $\alpha$ in the strong deflection limit, leading to an additional $1/b'$ dependence in~(\ref{eq:alpha0}) as well as different values for $\beta$, $\gamma$ and $\zeta$; details are provided in~\ref{sec:a=1}. 
  
\section{Asymptotic Approximant}\label{sec:asymptotics}
Although~(\ref{eq:alpha}) may be written in terms of elliptic integrals~\cite{Iyer:2009wa}, such integrals can only be evaluated numerically, making them computationally expensive to implement for the applications listed in Section~\ref{sec:intro}.  The series expansion for the weak limit~(\ref{eq:alpha1}) of light's bending angle may be used in lieu of elliptic integrals; however, this series converges slowly in its region of convergence, which extends from $b'=1$ towards (but not including) $b'=0$ (i.e., the radius of convergence is 1).   Thus, the strong limit can never be resolved by the series~(\ref{eq:alpha1}), regardless of how many terms are included, as the limit is nonuniform  in $b'$ (additional terms are needed to retain accuracy as $b'\to0$).

To overcome the issue of divergent or slowly converging series such as~(\ref{eq:alpha1}), there are a host of convergence acceleration (or ``re-summation'') techniques available that rely only on the original series itself (e.g. Pad\'e approximants, continued fractions,  Euler summation, etc.)~\cite{Bender}.  While such methods typically lead to an implementation improvement compared with the original series, global accuracy is not always guaranteed and the problem becomes one of choosing the ``best'' re-summation technique~\cite{Clisby,Guerrero,Tan}. However, for the current problem, additional information about the bending angle is available in the strong field limit, where divergence of the bending angle of light is described by~(\ref{eq:alpha0}).  These two disparate limits enable the use of asymptotic approximants~\cite{Barlow:2017} to construct an expression for the bending angle that is accurate from the weak to the strong limit.  One key feature, and advantage, of this technique is that it converges to the exact solution as additional terms are included from the series~(\ref{eq:alpha1}).

%A review of asymptotic approximants and their use is given in~\cite{Barlow:2017}.  The definition given in~\cite{Barlow:2017} of an asymptotic approximant is as follows:  
%\begin{definition}
%Given a power series representation of some function $f(x)$:
%\begin{equation}
%f=\sum_{n=0}^\infty a_n (x-x_0)^n,
%\label{powerseries}
%\end{equation}
%and an asymptotic behavior 
%\begin{equation}
%f\to f_a(x)~\mathrm{as}~x\to x_a,
%\label{asymp}
%\end{equation}
%an \textit{asymptotic approximant} is any function $f_A(x)$ that may be expressed analytically in closed form and that satisfies the following three properties:
%\begin{enumerate}
%\item The $N$-term Taylor expansion of $f_A$ about $x_0$ is identical to the $N$-term truncation of~(\ref{powerseries}).
%\item $\displaystyle{\lim_{x\to x_a} (f_A/f_a)=}$ constant for any $N$.
%%\item If it is known that $f$ is continuous on the open interval between $x_0$ and $x_a$,  $f_A$ should also be continuous on this interval.  
%\item The sequence of approximants converge for increasing $N$. 
%\end{enumerate}
%\label{defn}
%\end{definition}
%For the current problem $f(x)$ is $\alpha(b')$, $x_0$=1, $x_a$=0, and the two limiting behaviors are the weak and strong limits, given respectively by~(\ref{eq:alpha1}) and~(\ref{eq:alpha0}).  

\begin{figure*}[h!]
\begin{tabular}{cc}(a)\hspace{-.05in}
\includegraphics[width=2.9in]{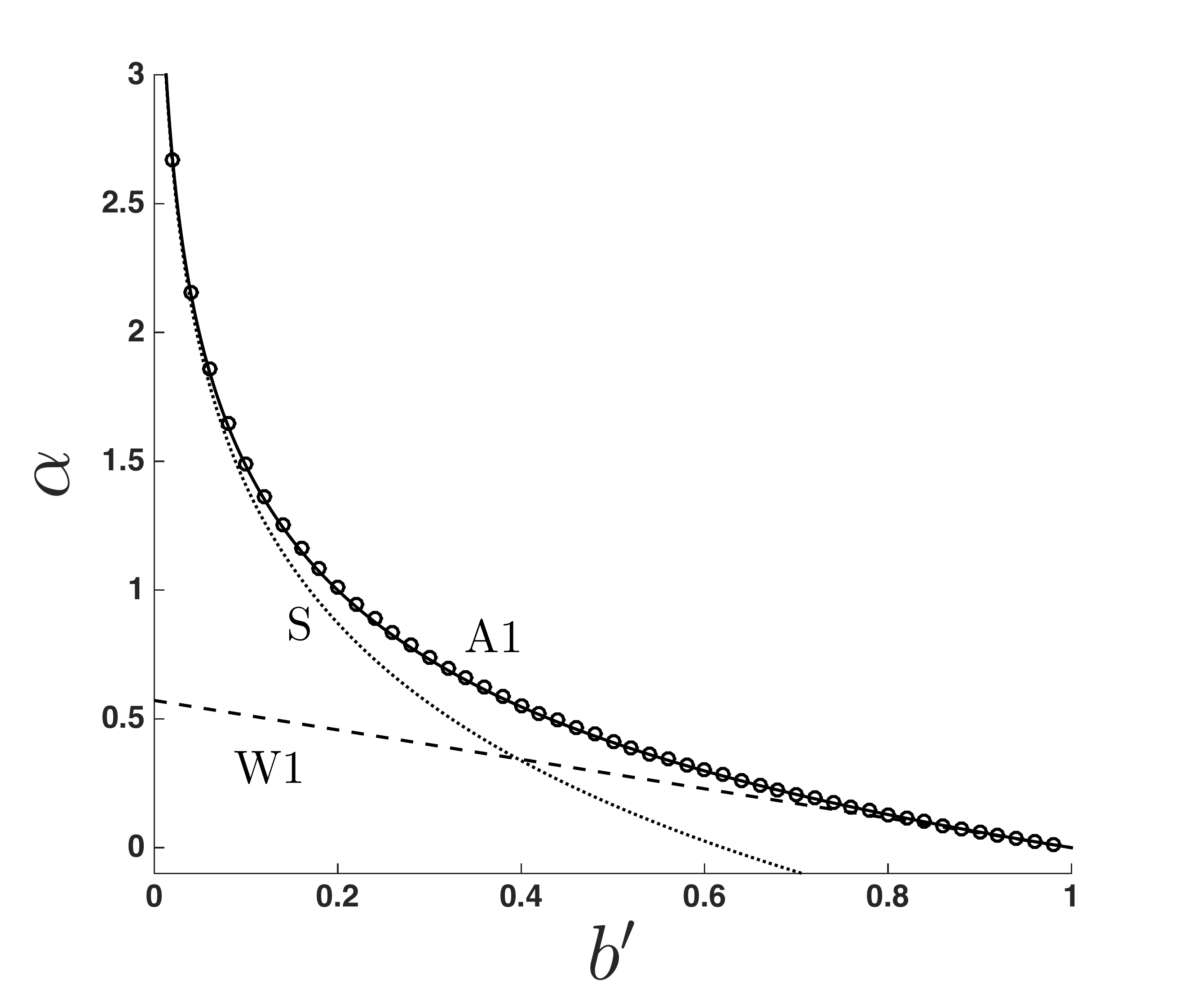}
(b)\hspace{-.05in}
\includegraphics[width=2.9in]{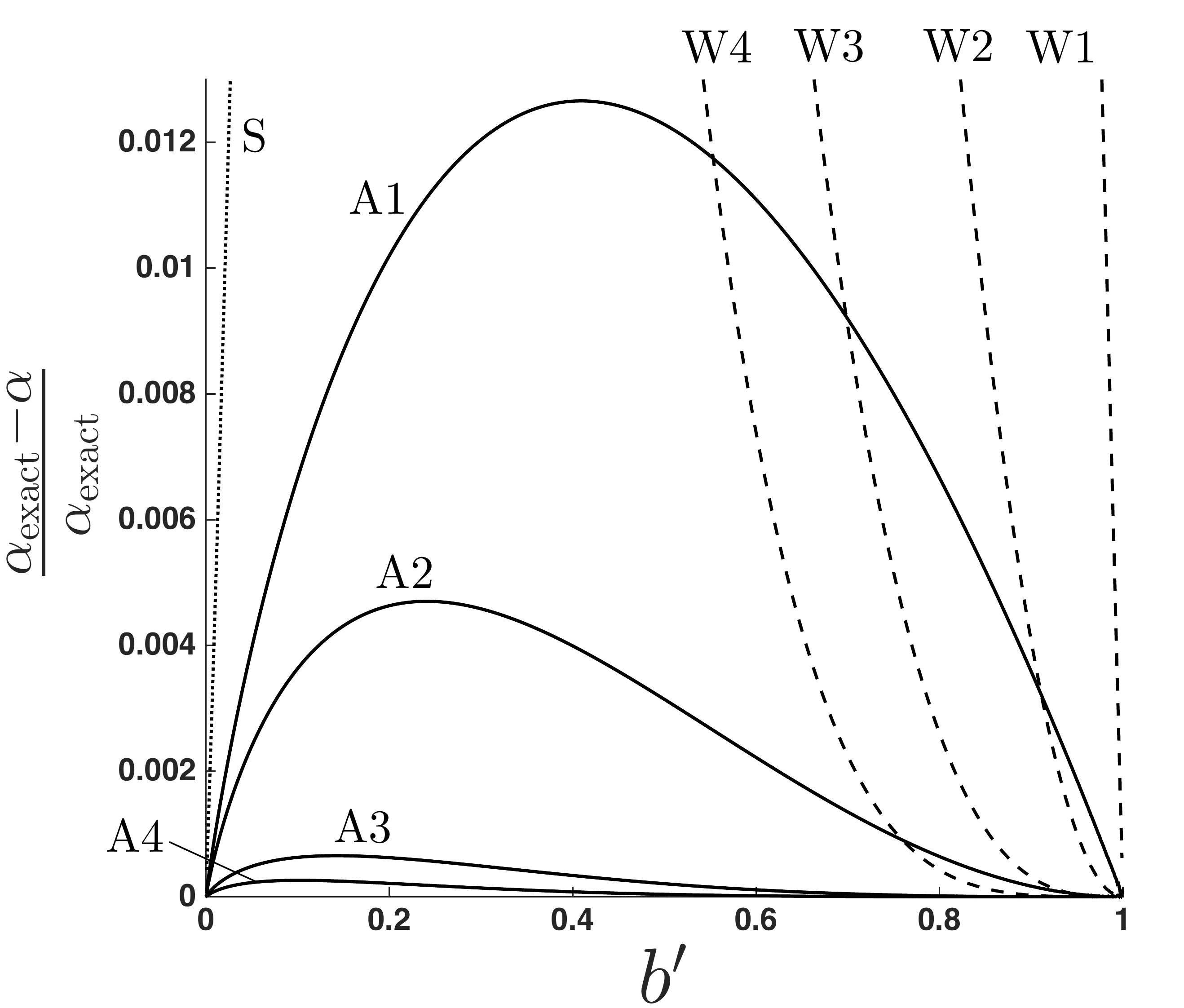}
\end{tabular}
\caption{Bending angle of light, $\alpha$, as a function of the impact perturbation parameter, $b'$~(\protect\ref{eq:impact}), for an $a=-1$ Kerr black hole.  Comparison between the $N$-term weak deflection limit Kerr series~(\protect\ref{eq:alpha1}) [dashed curves] labeled as W$N$, the corresponding approximant~(\protect\ref{eq:approximant}) labeled as A$N$ [solid curves], and the leading-order strong deflection limit~(\protect\ref{eq:alpha0}) [dotted curve] labeled as S. (a) comparison with ``exact'' (converged numerical) solution ($\circ$) of~(\protect\ref{eq:alpha}). (b)  relative error of the approximant.}
\label{fig:aminus1}
\end{figure*}

\begin{figure*}[h!]
\begin{tabular}{cc}(a)\hspace{-.05in}
\includegraphics[width=2.9in]{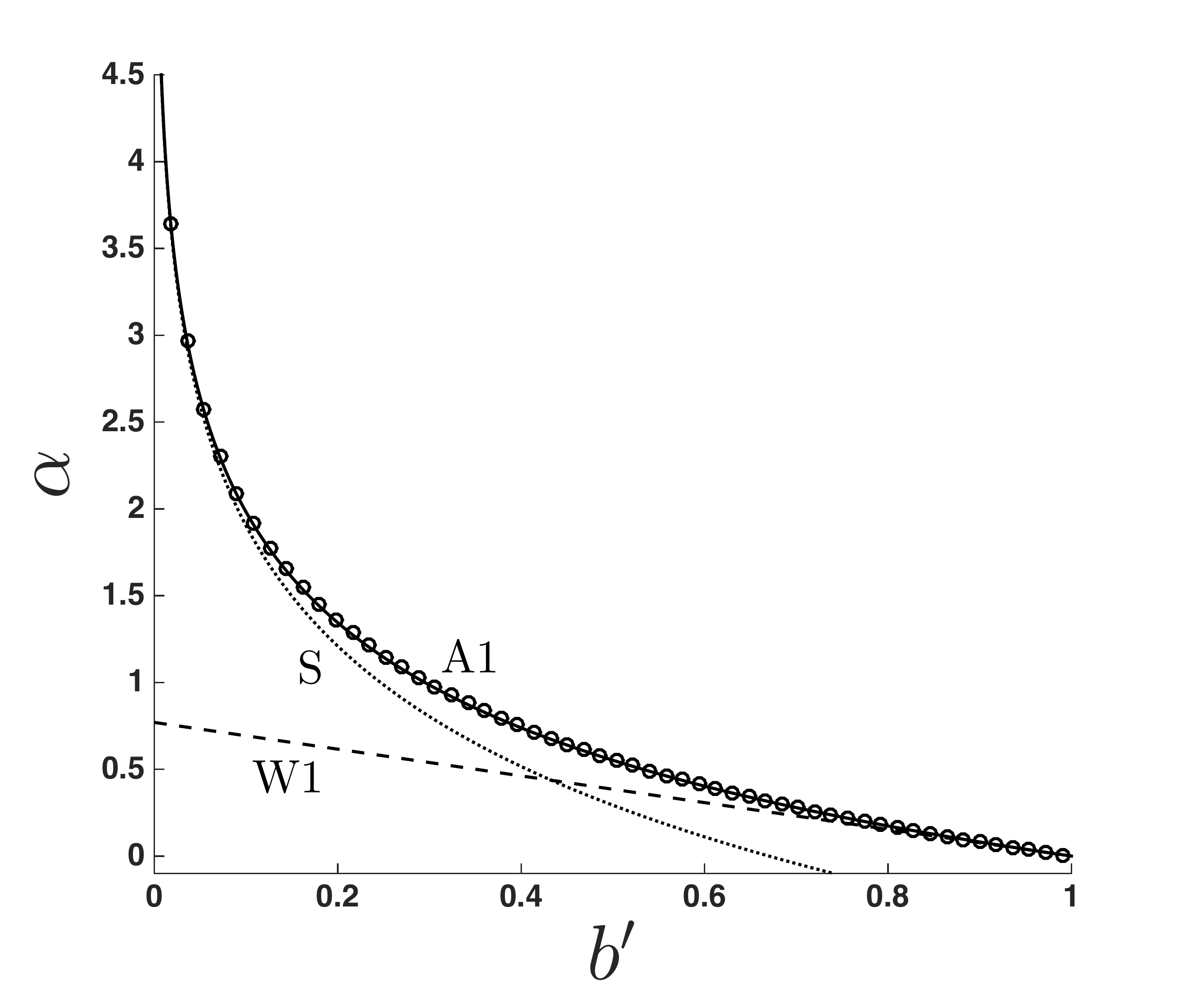}
(b)\hspace{-.05in}
\includegraphics[width=2.9in]{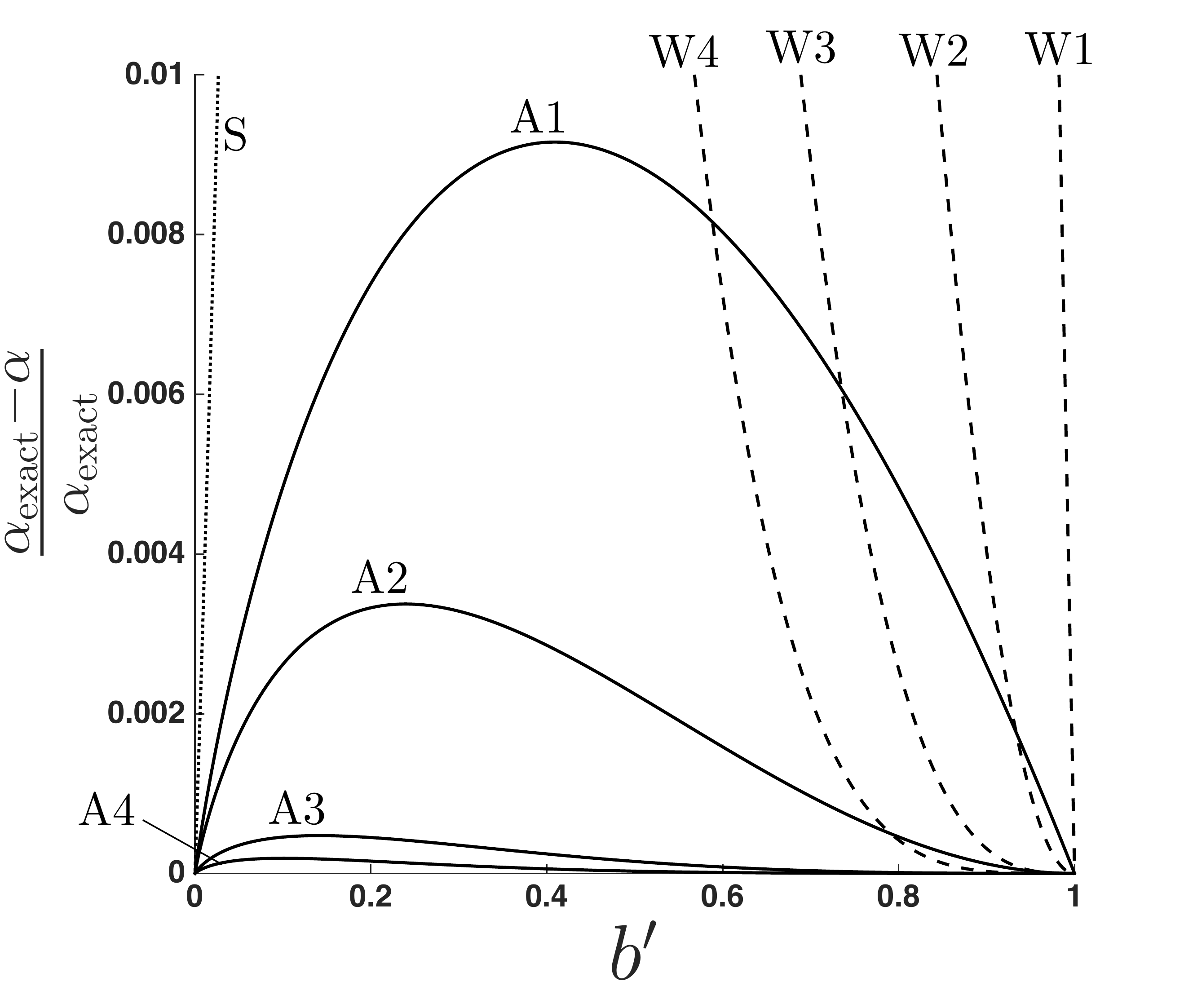}
\end{tabular}
\caption{Bending angle of light, $\alpha$, as a function of the impact perturbation parameter, $b'$~(\protect\ref{eq:impact}), for a Schwarzschild black hole ($a=0$).   Conventions are as in figure~\protect\ref{fig:aminus1}. (a) approximant~(\protect\ref{eq:approximant}) compared with ``exact'' (converged numerical) solution of~(\protect\ref{eq:alpha}). (b)  relative error of the approximant.}
\label{fig:a0}
\end{figure*}

\begin{figure*}[h!]
\begin{tabular}{cc}(a)\hspace{-.05in}
\includegraphics[width=2.9in]{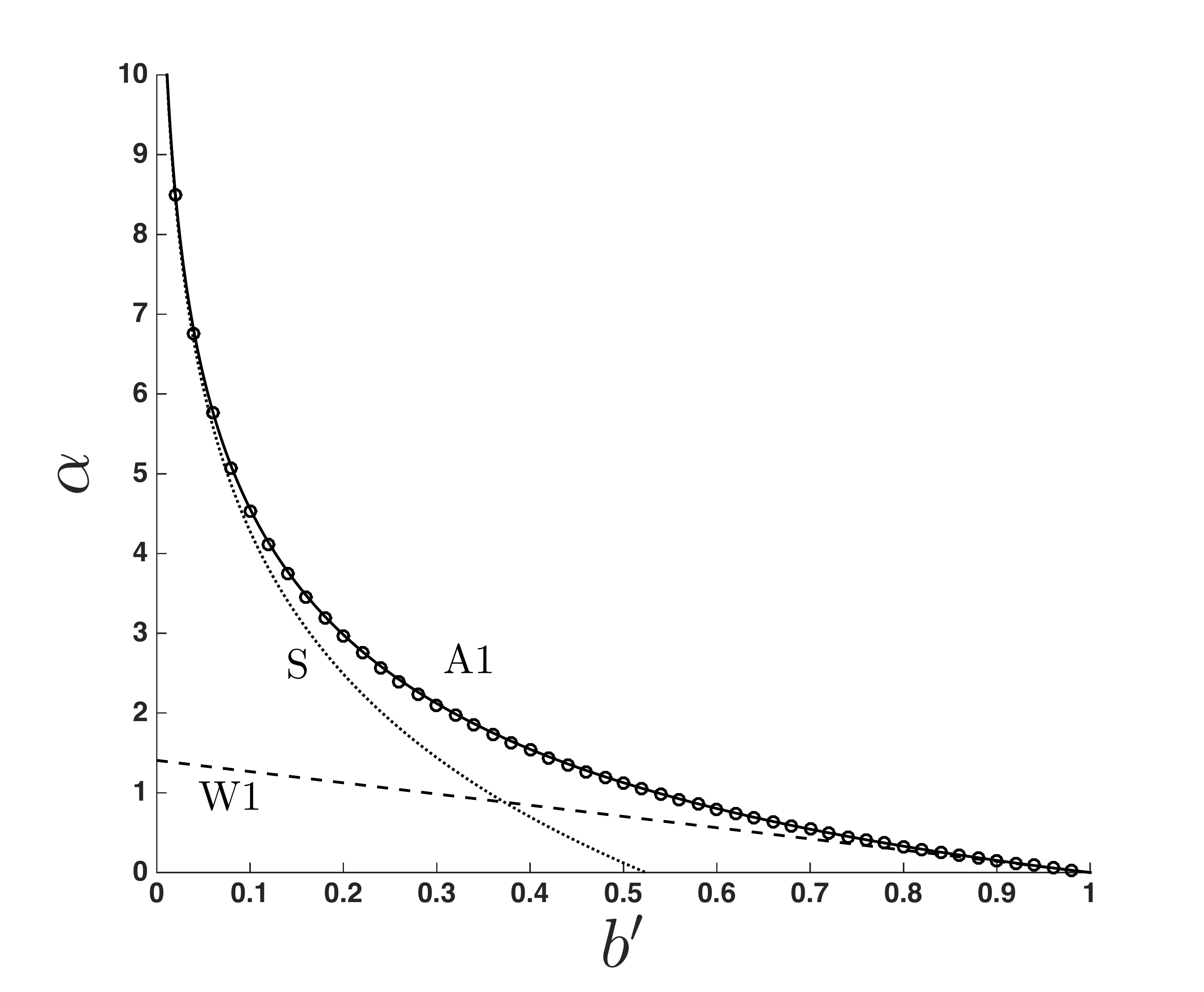}
(b)\hspace{-.05in}
\includegraphics[width=2.9in]{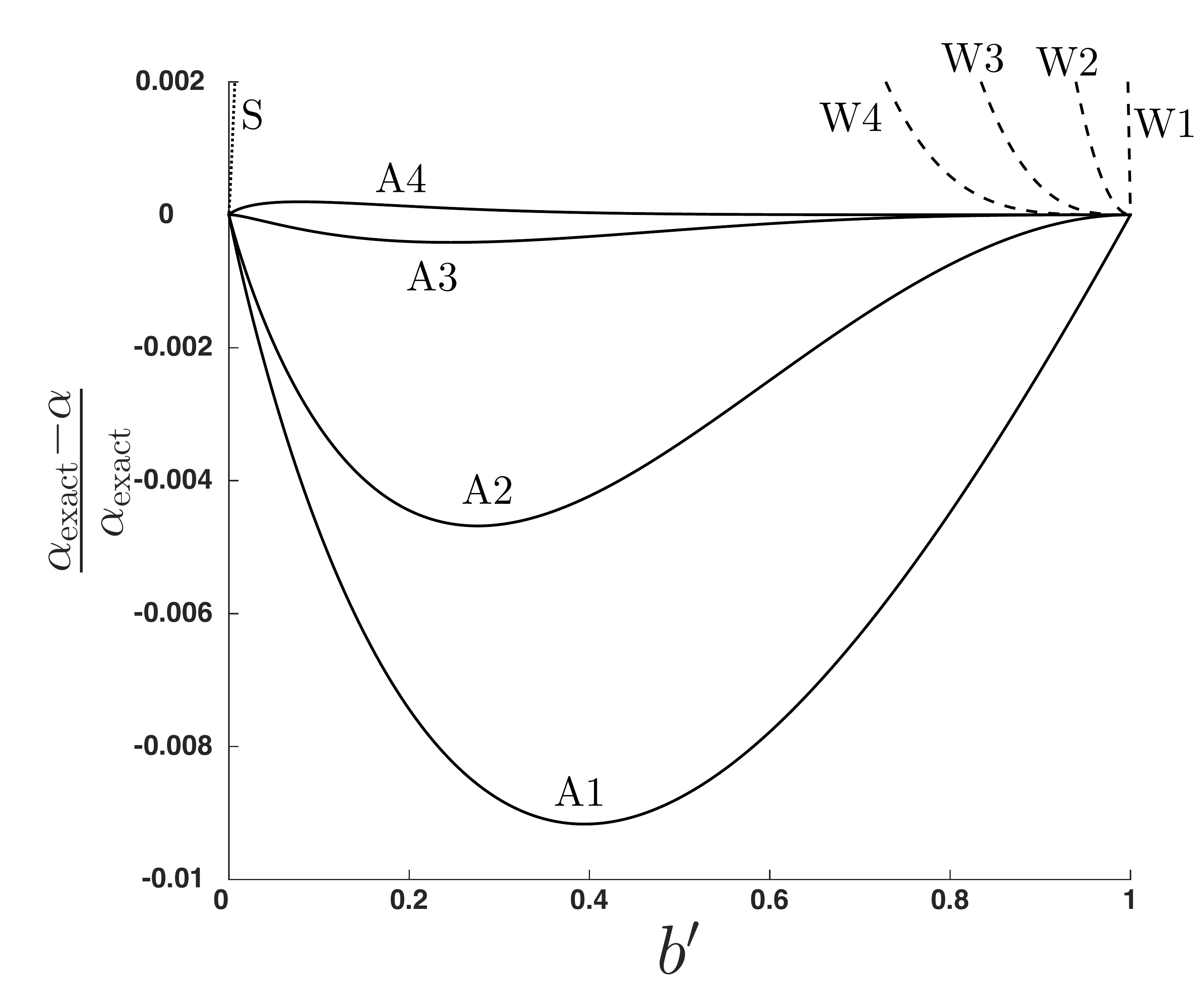}
\end{tabular}
\caption{Bending angle of light, $\alpha$, as a function of the impact perturbation parameter, $b'$~(\protect\ref{eq:impact}), for an $a=0.9$ Kerr black hole.   Conventions are as in figure~\protect\ref{fig:aminus1}. (a) approximant~(\protect\ref{eq:approximant}) compared with ``exact'' (converged numerical) solution of~(\protect\ref{eq:alpha}). (b)  relative error of the approximant.}
\label{fig:apt9}
\end{figure*}

\begin{figure*}[h!]
\begin{tabular}{cc}(a)\hspace{-.05in}
\includegraphics[width=2.9in]{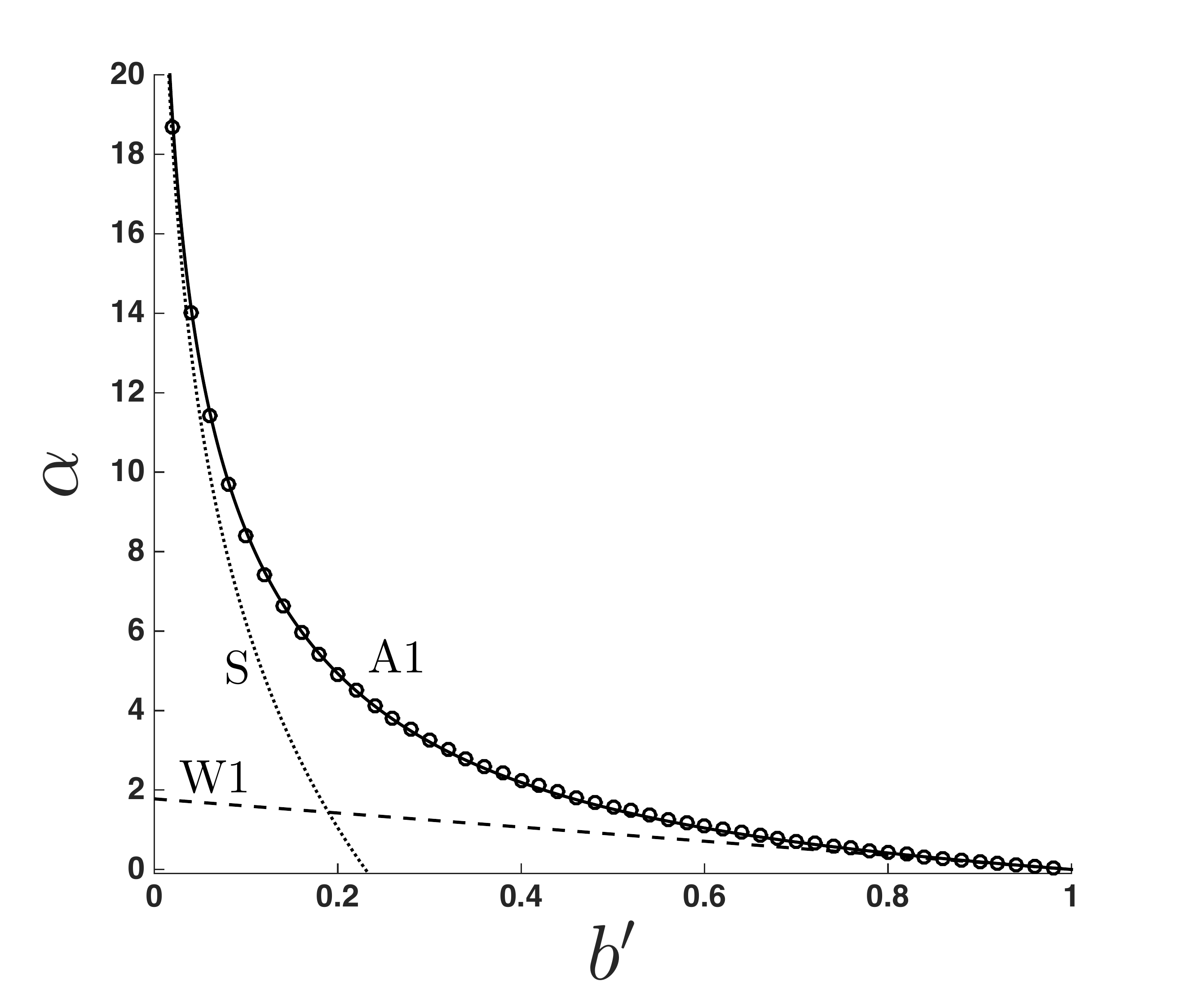}
(b)\hspace{-.05in}
\includegraphics[width=2.9in]{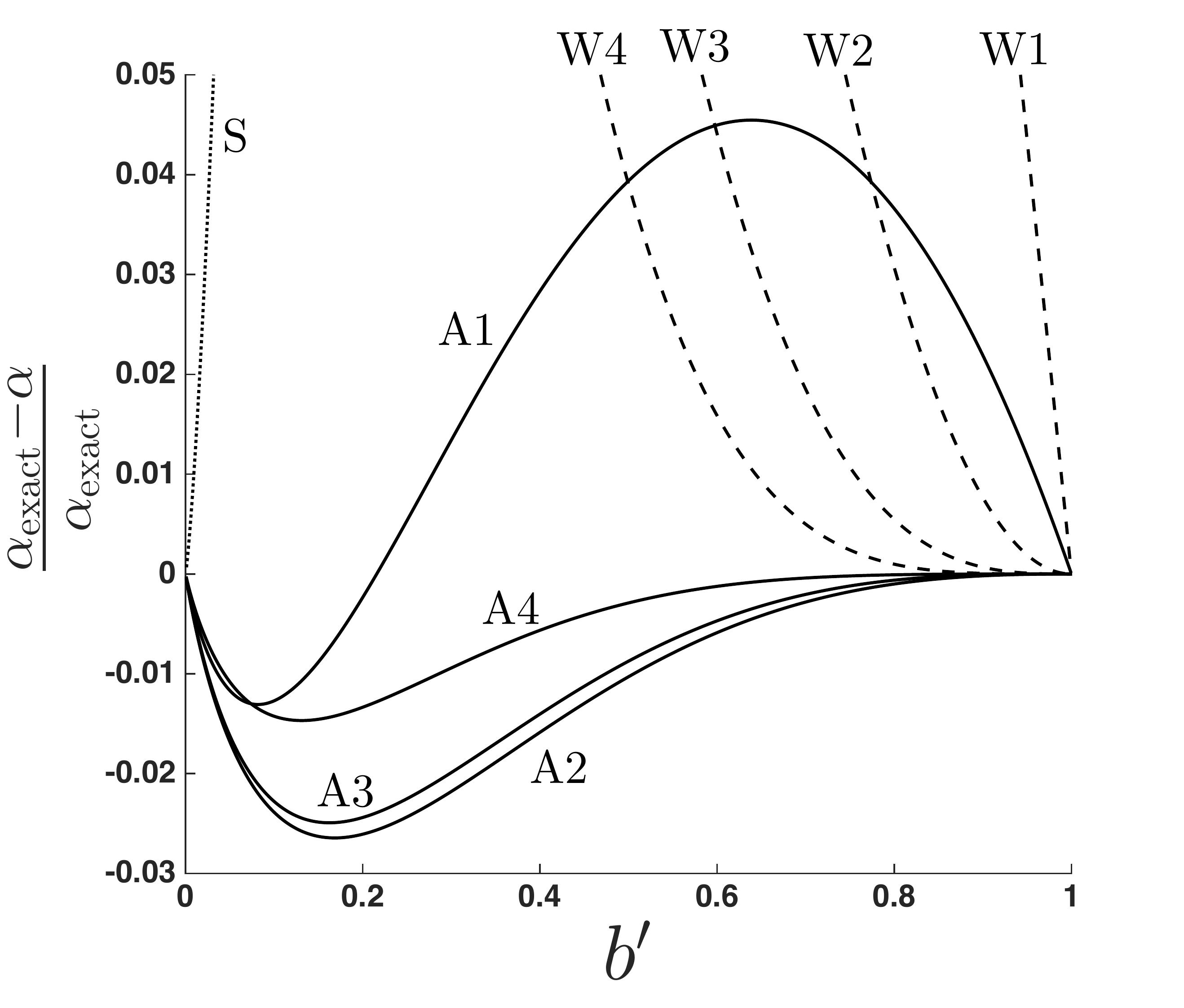}
\end{tabular}
\caption{Bending angle of light, $\alpha$, as a function of the impact perturbation parameter, $b'$~(\protect\ref{eq:impact}), for an $a=0.99$ Kerr black hole.   Conventions are as in figure~\protect\ref{fig:aminus1}. (a) approximant~(\protect\ref{eq:approximant}) compared with ``exact'' (converged numerical) solution of~(\protect\ref{eq:alpha}). (b)  relative error of the approximant.}
\label{fig:apt99}
\end{figure*}

\begin{figure*}[h!]
\begin{tabular}{cc}(a)\hspace{-.05in}
\includegraphics[width=2.9in]{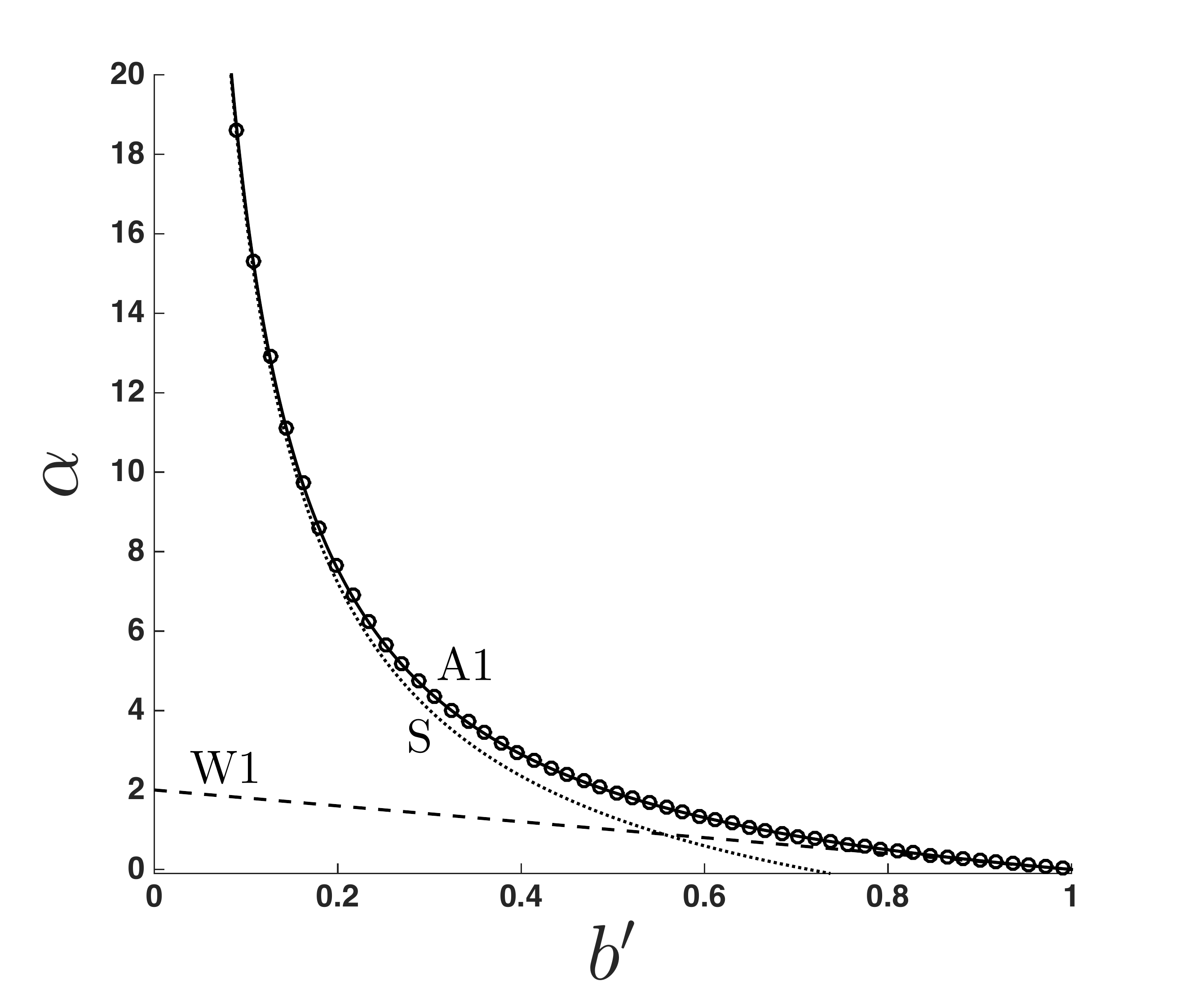}
(b)\hspace{-.05in}
\includegraphics[width=2.9in]{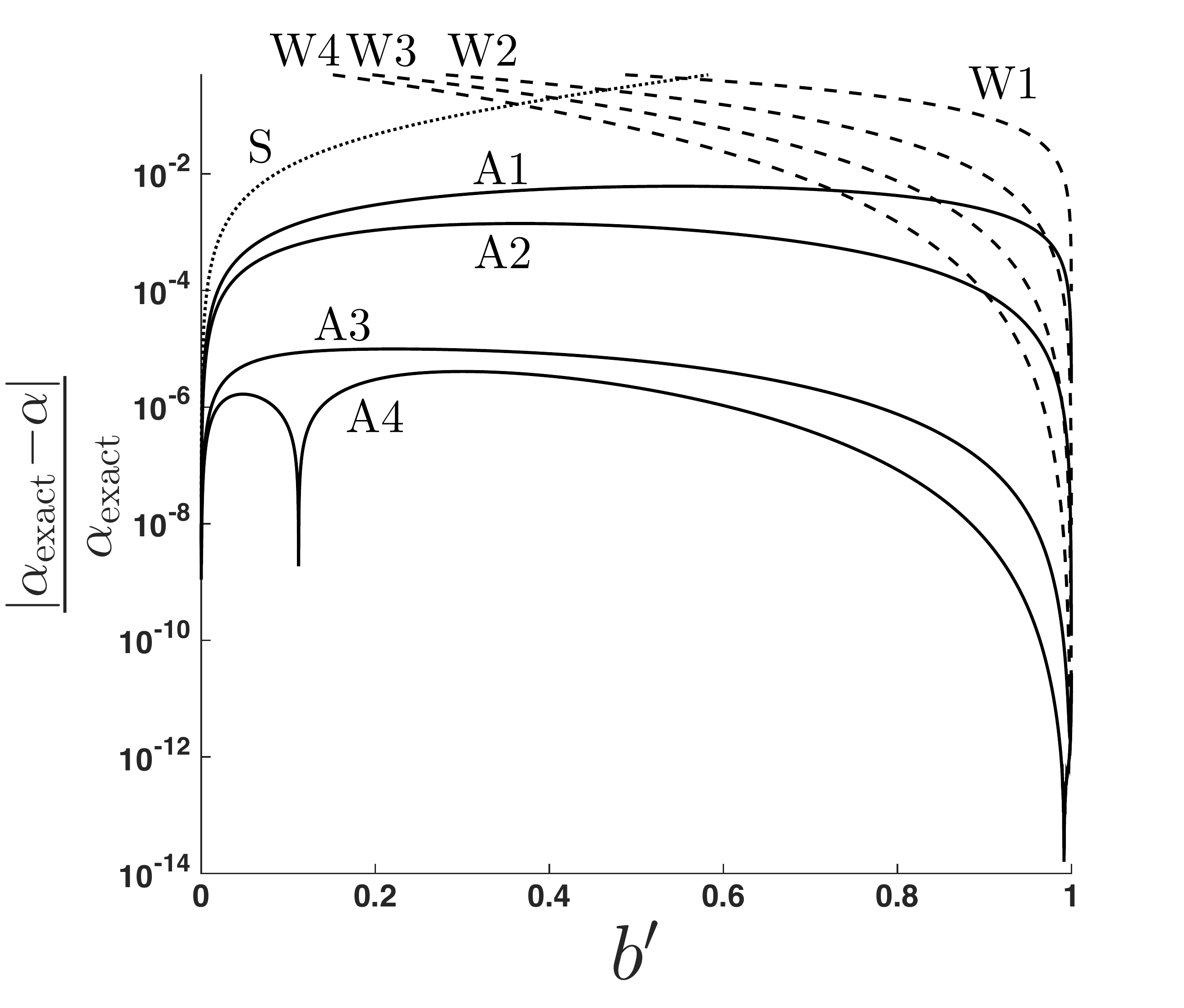}
\end{tabular}
\caption{Bending angle of light, $\alpha$, as a function of the impact perturbation parameter, $b'$~(\protect\ref{eq:impact}), for an extremal Kerr black hole ($a=1$, prograde orbit).    Conventions are as in figure~\protect\ref{fig:aminus1}. (a) approximant~(\protect\ref{eq:approximant}) compared with ``exact'' (converged numerical) solution of~(\protect\ref{eq:alpha}). (b)  relative error of the approximant.}
\label{fig:a1}
\end{figure*}

The form of the approximant is chosen such that it limits to~(\ref{eq:alpha0}) as $b'\to0$.  As shown in~\cite{Iyer:2006cn} (for $a=0$) and~\ref{sec:a=1} (for $a=1$), a full asymptotic expansion of~(\ref{eq:alpha}) about $b'\to0$ invokes higher-order terms (i.e., beyond those in~(\ref{eq:alpha0})) of orders $b'\ln b'$, $b'$, $b'^2\ln b'$, $b'^2$, etc.  These terms are analytic at $b'=1$, and thus can be incorporated as auxiliary terms in an approximant that bridges these two limits, as similarly done in~\cite{Barlow:2017} for the Sakiadis boundary layer problem.  An asymptotic approximant for describing light deflection in the Kerr black hole may be expressed as
\begin{equation}
 \alpha_{{\rm A}N}=-\pi+\beta+\gamma\ln\zeta+\delta_{a,1}\frac{\sqrt{3}}{b'}-\gamma\ln b'+\sum_{n=1}^{N+1}B_nb'^{\frac{n}{2}}\left(\Delta_{n+1}\sqrt{b'}\ln b'+\Delta_n\right)
 \label{eq:approximant}
\end{equation}
where $\Delta_n=1+(-1)^n$.  Approximant $\alpha_{{\rm A}N}$ (given by ~(\ref{eq:approximant})) matches the leading order asymptotic behavior of $\alpha$ as $b'\to0$ (given by~(\ref{eq:alpha0})) and the $B_n$ coefficients are computed such that  the expansions of $\alpha_{{\rm A}N}$ and $\alpha$ about $b'=1$ (given by~(\ref{eq:alpha1})) are identical to $N^\mathrm{th}$-order.   The latter is accomplished by setting the series in~(\ref{eq:alpha1}) equal to the expansion of~(\ref{eq:approximant}) about $b'=1$ and then equating like terms in $(b'-1)^0$, $(b'-1)^1$, $(b'-1)^2$, \dots, $(b'-1)^N$; this leads to the following linear system of $N+1$ equations: 
\begin{eqnarray}
\nonumber
D_0&=&\sum_{n=1}^{N+1}B_n\Delta_n\\
\nonumber
D_1&=&\sum_{n=1}^{N+1}B_n\left(\Delta_{n+1}+\frac{\Delta_n}{1!}\frac{n}{2}\right)\\
\nonumber
D_2&=&\sum_{n=1}^{N+1}B_n\left[\Delta_{n+1}\frac{n}{2}+\frac{\Delta_n}{2!}\frac{n}{2}\left(\frac{n}{2}-1\right)\right]\\
\nonumber
&\vdots&\\
\nonumber
D_j&=&\sum_{n=1}^{N+1}B_n\left[\Delta_{n+1}C_{n,j}+\frac{\Delta_n}{j!}\prod_{k=0}^{j-1}\left(\frac{n}{2}-k\right)\right]\\
\nonumber
&\vdots&\\
D_N&=&\sum_{n=1}^{N+1}B_n\left[\Delta_{n+1}C_{n,N}+\frac{\Delta_n}{N!}\prod_{k=0}^{N-1}\left(\frac{n}{2}-k\right)\right]
\label{matrix}
\end{eqnarray}
where
\begin{equation}
D_{j>0}=a_j+(-1)^j\left(\frac{\gamma}{j}-\delta_{a,1}\sqrt{3}\right),~D_0=\pi-\beta-\gamma\ln\zeta-\delta_{a,1}\sqrt{3}
\label{D}
\end{equation}
and
\begin{equation}
C_{n,j}=\frac{(-1)^{j+1}}{j}-\sum_{m=1}^{j-1}\left[\frac{(-1)^{j-m}}{m!(j-m)}\prod_{k=0}^{m-1}\left(\frac{n+1}{2}+k\right)\right].
\label{product}
\label{A}
\end{equation}
Written explicitly for $N=4$,~(\ref{matrix}) becomes
\begin{equation}\left[\begin{tabular}{l}
\hline
\multicolumn{1}{|l|}{$D_0$} \\ 
\multicolumn{1}{|l|}{$D_1$} \\ \hline
$D_2$                       \\
$D_3$                       \\
$D_4$                      
\end{tabular}     \right]=\left[
\begin{tabular}{ccccc}
\cline{1-2}
\multicolumn{1}{|c}{0} & \multicolumn{1}{c|}{2} & 0      & 2 & 0    \\ 
\multicolumn{1}{|c}{2} & \multicolumn{1}{c|}{2} & 2      & 4 & 2    \\ \cline{1-2}
1                       & 0                      & 3      & 2 & 5    \\
$-$1/3                  & 0                      & 2/3    & 0 & 11/3 \\
1/6                     & 0                      & $-$1/6 & 0 & 1/2 
\end{tabular}\right]\left[\begin{tabular}{l}
\hline
\multicolumn{1}{|l|}{$B_1$} \\ 
\multicolumn{1}{|l|}{$B_2$} \\ \hline
$B_3$                       \\
$B_4$                       \\
$B_5$                      
\end{tabular}     \right].
\label{eq:A5}
\end{equation}
Note that, since the elements of~(\ref{matrix}) are independent of $N$,~(\ref{eq:A5}) also includes the coefficient matrices for all lower-order approximants.  The simplest form of the approximant is for $N=1$, indicated by the boxed system in~(\ref{eq:A5}).  Solving this system, the $N=1$ approximant is given by
\begin{eqnarray}
\nonumber
\alpha_\mathrm{A1}=&\left(-\pi+\beta+\gamma\ln\zeta\right)\left(1-b'+b'\ln b'\right)-4\left(b'\ln b'\right)/b_c\\
&+\gamma\left(b'-1\right)\ln b'+\delta_{a,1}\sqrt{3}\left(1/b'-b'+2b'\ln b'\right)
\end{eqnarray}
and is shown in figures~\ref{fig:aminus1} through~\ref{fig:a1} (labeled as A1), compared with the numerical solution of~(\ref{eq:alpha}) for different values of spin $a$.  Note, that although $\alpha_\mathrm{A1}$ is constructed using only a first-order truncation of the $b'=1$ expansion~(\ref{eq:alpha1}) (labeled W1 in figures) and the leading order $b\to0$ behavior given by~(\ref{eq:alpha0}) (labeled S in figures), accuracy is preserved between these two limits.  For additional accuracy, one may invert the larger sub-matrices in~(\ref{eq:A5}) to obtain the $N$=2, 3, and 4 approximants, also shown in the figures (labeled A2, A3, A4).  For convenience, the $N=4$ approximant, $\alpha_\mathrm{A4}$, is given by~(\ref{eq:approximant}) with coefficients
\begin{eqnarray}
\nonumber
B_1=-\frac{17}{6}\left(D_0-D_1+D_2\right)+3D_3-5D_4\\
\nonumber
B_2=-4D_0+\frac{9}{2}D_1-5D_2+6D_3-12D_4\\
\nonumber
B_3=-\frac{7}{3}\left(D_0-D_1+D_2\right)+3D_3-8D_4\\
\nonumber
B_4=\frac{9}{2}\left(D_0-D_1\right)+5D_2-6D_3+12D_4\\
B_5=\frac{1}{6}\left(D_0-D_1+D_2\right)+D_4.
\end{eqnarray}

\begin{figure*}[h!]
\begin{tabular}{cc}(a)\hspace{-.05in}
\includegraphics[width=2.9in]{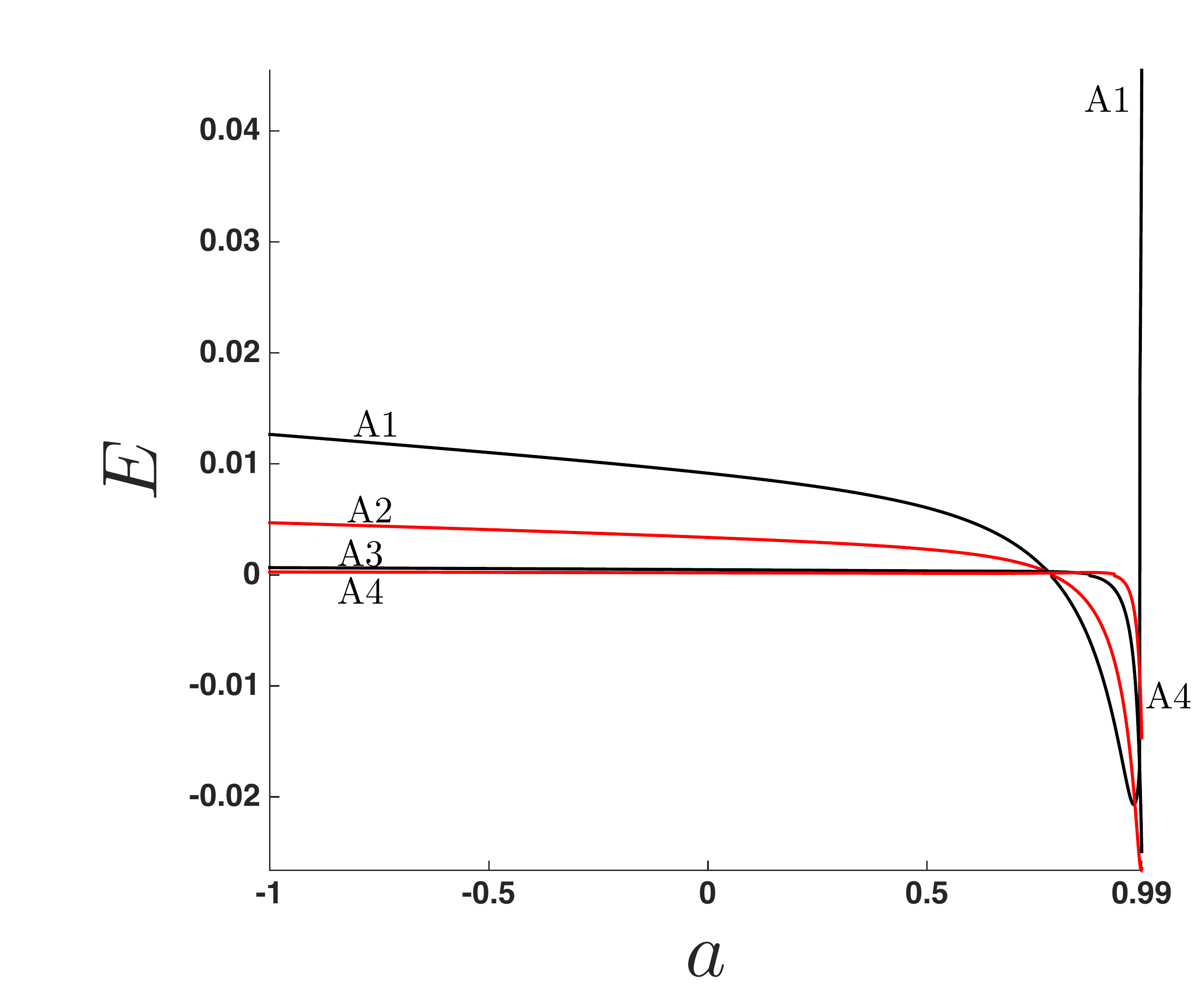}
(b)\hspace{-.05in}
\includegraphics[width=2.9in]{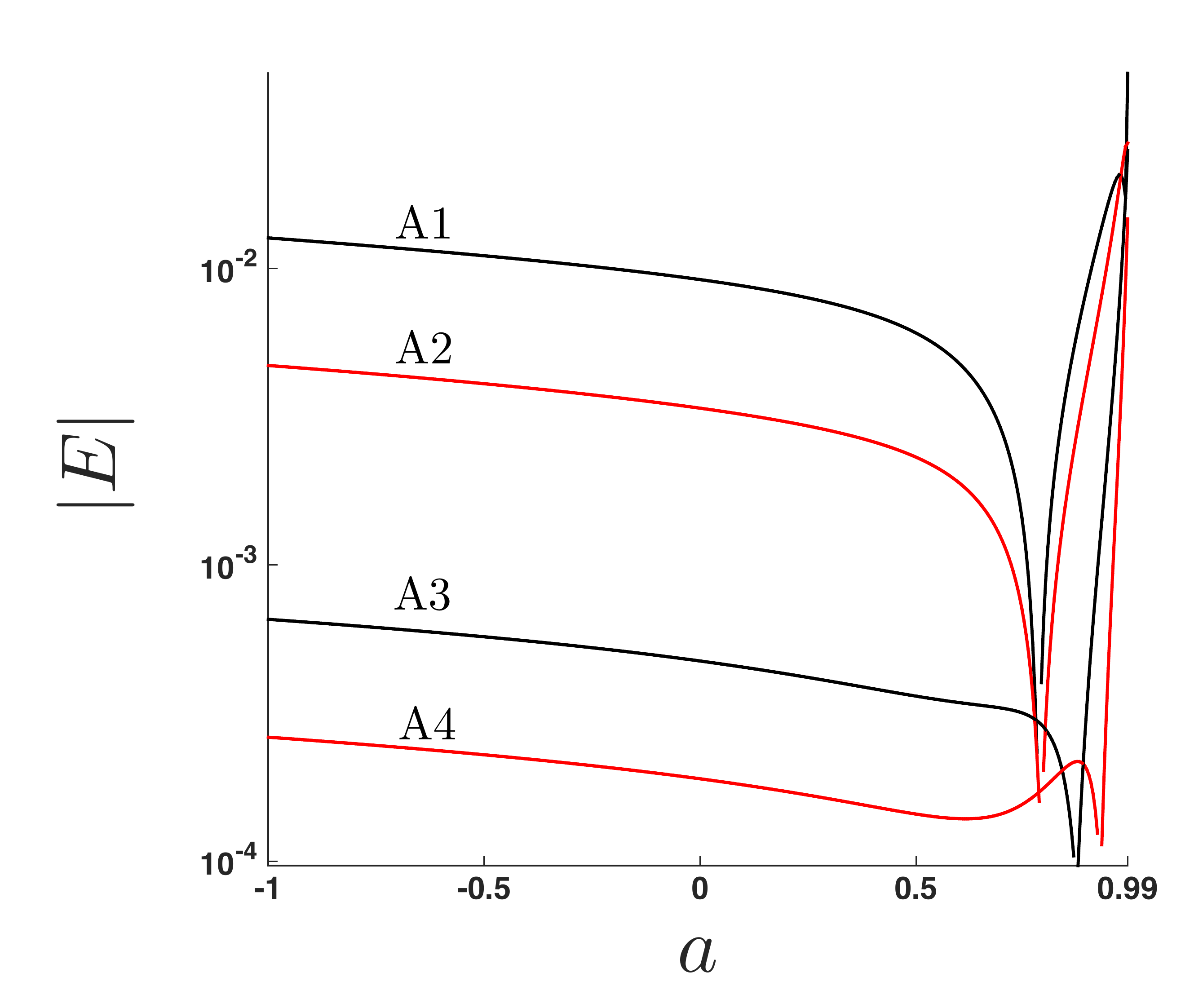}
\end{tabular}
\caption{(a) Extremum error, $E$, defined as the global extremum (maximum or minimum) of the normalized relative error $\left(\alpha_\mathrm{exact}-\alpha_{{\rm A}N}\right)/ \alpha_\mathrm{exact}$ in the approximant~(\ref{eq:approximant}) (labeled as A$N$) over the interval $0\le b'\le1$, shown for $a<1$. (b) $|E|$ (infinity norm of normalized relative error), shown on a semi-log plot for clarity. Note that for $a=1$ (not shown on the figure), the infinity norm of A4 is $O(10^{-6})$ (see figure~\ref{fig:a1}) due to the special treatment of the $a=1$ strong deflection limit incorporated in approximant~(\ref{eq:approximant}). }
\label{fig:error}
\end{figure*}

The distributions of normalized relative error, $\left(\alpha_\mathrm{exact}-\alpha_{{\rm A}N}\right)/ \alpha_\mathrm{exact}$,  for $a=-1$ (figure~\ref{fig:aminus1}b) and $a=0$ (Schwarzschild case, figure~\ref{fig:a0}b) are nearly identical.  This uniformly convergent behavior over the full interval $0\le b'\le1$ holds for nearly all spins in the range $-1\le a\lesssim0.99$.  Plots of the error (versus $b'$) in this range (not shown) have similar shapes to that shown in figures~\ref{fig:aminus1}b,~\ref{fig:a0}b, and~\ref{fig:apt9}b, except with slightly different magnitudes and $b'$ locations for the global maxima and/or minima.  Figure~\ref{fig:error}a indicates that the global extrema of the error (for $-1\le a\lesssim0.99$) decreases as $N$ increases, except in the vicinity of the region where the error changes sign; even in this region, the infinity norm of the error is at most $O(10^{-2})$, as indicated by the curves surrounding the discontinuities in~figure~\ref{fig:error}b.

% While this convergence behavior ultimately continues for larger $a\lesssim0.9$, it is not true for all $a$ values within the range $0.75\lesssim a\lesssim0.9$, as indicated by the intersection of curves surrounding the discontinuities in figure~\ref{fig:error}a, at various orders $N$.  The discontinuities in figure~\ref{fig:error}a are an indication of the approximant transitioning from an under-approximation to an over-approximation of the exact solution, as indicated in figure~\ref{fig:error}b, which shows the global extremum error (maximum or minimum) over the full range of $a$.   While this transition from a global maximum error to a global minimum error in the range $0.75\lesssim a\lesssim0.9$ leads to a non-uniformity of accuracy across $0\le b'\le1$ for certain $a$ values in this range, the maximum relative error is at most $O(10^{-2})$, as indicated in figure~\ref{fig:error}a.  

%The global maximum is tracked   This is shown in figure~\ref{fig:error} where, for $a\lesssim0.75$, global maxima is tracked for error curves that are identical in shape to those given in figure~\ref{fig:aminus1}b. 

Convergence of the approximant becomes nonuniform as $a\to1$, as can be seen by comparing~figures~\ref{fig:aminus1} through~\ref{fig:apt99}.    In figures~\ref{fig:aminus1}a through~\ref{fig:apt99}a, there is a range of $b'$ for which neither the leading-order strong nor weak deflection limits (denoted by S and W1, respectively) agree with the exact solution.  The approximant bridges these limits, but requires additional terms for $a=0.99$ (to retain the same accuracy) because the range of $b'$ between accurate asymptotic representations is larger than for  $a=-1$, 0, and 0.9.  For example, at the location $b'=0.5$ in figures~\ref{fig:aminus1}b through~\ref{fig:apt9}b (for $a=-1$, 0, and 0.9), the error in approximant A1 is $<0.02$. To achieve this same level of accuracy for $a=0.99$ at $b'=0.5$, an additional term is required in the approximant  (i.e, the A2 curve), as shown in figure~\ref{fig:apt99}b.  This convergence behavior can be explained by the non-uniform limit in~(\ref{eq:alpha0}) as $a\to1$. The effect of this is shown in figure~\ref{fig:error} as a dramatic  jump in error near $a=0.99$.  Although the leading-order divergent behavior as $b'\to0$ in~(\ref{eq:alpha0}) is logarithmic for $a\neq1$, this divergence becomes subdominant to the $\sqrt{3}/b'$ term when $a=1$.   As $a\to1$, the logarithmic behavior  (while still responsible for divergence) occurs for for an increasingly tighter region near $b'\to0$, before invoking a $\sqrt{3}/b'$ behavior when $a=1$.  The effect of this is that additional terms in~(\ref{eq:alpha1}) and (by-extension) the approximant~(\ref{eq:approximant}) are required to maintain accuracy as $a\to1$ (but $a\neq1$) and as $b'\to0$.  For $a=1$, the approximant simply enforces the leading-order $\sqrt{3}/b'$ term in addition to the subdominant logarithmic behaviors, leading to the  approximant shown in figure~\ref{fig:a1} with correspondingly small errors.

\section{Conclusions}\label{sec:conclusions}
In this work, we develop closed-form expressions for photon deflection along the equatorial plane for Kerr black holes of any spin, using the method of asymptotic approximants.  The approximants are designed so that the strong and weak deflection limits may be joined uniformly.  As additional terms are included, the approximants converge to the exact solution across the full range of impact parameter.  As elliptical integrals are computationally expensive, the high accuracy of these asymptotic approximants make them an attractive alternative for use in black hole simulations.  We have also derived an expression for the strong deflection limit of a prograde orbit around an extremal ($a=1$) Kerr black hole, including inverse power, logarithmic, and constant terms; to the authors' knowledge, this is the first time the full non-vanishing behavior of this limit has been determined.

This work supports the recent development of asymptotic approximants as a useful tool in mathematical physics.   While the accuracy we have found for the cases considered here is quite high, we also note that a key benefit of our method is the uniformity of the convergence. In particular, including a sufficient number of  terms in the expansion  allows for  arbitrary accuracies to be obtained. Finally, we note that the results of this paper augment those presented in the method paper of Barlow et al.~\cite{Barlow:2017}, and demonstrate the increasing number of problems in mathematical physics that are well-described by  asymptotic approximants. 
 
 \ack{The authors wish to thank Richard O'Shaughnessy and Yosef Zlochower for helpful conversations.  JAF was supported by NSF award ACI-1550436.}

% \clearpage
\section*{References}
%\begin{thebibliography}{99}
\bibliography{asymptotic}
%\end{thebibliography}

\appendix
\section{The strong deflection limit for a prograde orbit around an extremal ($a=1$) Kerr black hole}\label{sec:a=1}

The analysis in~\cite{Iyer:2009hq} breaks down for the case of extremal Kerr BHs ($a=1$) in the strong-deflection regime for prograde orbits.  There are important physical reasons for this, particularly the fact that the Innermost Circular Orbit in this case  coincides with the BH event horizon at coordinate radius $r=1$.  The goal of this appendix is to determine the asymptotic behavior of the deflection angle in the strong deflection limit for $a=1$.

We may simplify~(\ref{eq:alpha}) for $a$=1 by identifying $r_0=b-1$, leading to
\begin{eqnarray}
\alpha +\pi&=& 2\int_0^{u_0} \frac{1-2u(1-1/b)}{(1-u)^2\sqrt{\frac{2(b-1)^2}{b^2}\left(u^3-\frac{b+1}{2(b-1)}u^2+\frac{1}{2(b-1)^2}\right)}}\, du.\nonumber\\
&=&\sqrt{2} \int_0^{u_0}\frac{1+u_0-2u}{(1-u)^2\sqrt{u^3-\left(\frac{1+2u_0}{2}\right)u^2+\frac{u_0^2}{2}}}\,du.
\label{eq:maximal}
\end{eqnarray}
We are interested in the behavior when $u_0\sim 1$, so we may define $\epsilon\equiv 1-u_0$ and examine the behavior as $\epsilon\to0$.  Making the above substitution and the additional substitution $Y=\sqrt{\frac{1-\epsilon-u}{\epsilon-\epsilon^2}}$ for the integration variable,~(\ref{eq:maximal}) becomes
\begin{equation}
\alpha +\pi=\frac{2^{3/2}}{\epsilon}\int_0^{\epsilon^{-1/2}}\!\!\frac{1+2Y^2-2\epsilon Y^2}{(1+Y^2-\epsilon Y^2)^2\sqrt{1+3Y^2/2-\epsilon(Y^4+2Y^2)+\epsilon^2Y^4}}\,dY.
\label{eq:alphaY}
\end{equation}
Note that the above integral diverges as $\epsilon\to0$, and our goal is to determine precisely how it diverges, i.e., the asymptotic behavior in this limit.  To improve the accuracy of an approximant derived from the asymptotic behavior, we wish to determine all non-vanishing terms in the limit. These details inform the choice of asymptotic approximant used in Section~\ref{sec:asymptotics}. 

 Expanding the integrand of~(\ref{eq:alphaY}) about $\epsilon=0$ leads to
\begin{eqnarray}
\nonumber
\alpha+\pi&=\frac{1}{\epsilon}\int_0^{\epsilon^{-1/2}}\frac{8Y^2+4}{(1+Y^2)^2\sqrt{2+3Y^2}}\,dY+\int_0^{\epsilon^{-1/2}}\frac{8Y^2+52Y^6+44Y^4+8Y^2}{(1+Y^2)^3(2+3Y^2)^{3/2}}\,dY\\
\nonumber
&+ \epsilon\int_0^{\epsilon^{-1/2}}\frac{2Y^6(6Y^8+39Y^6+138Y^4+145Y^2+46)}{(1+Y^2)^4(2+3Y^2)^{5/2}}\,dY\\&+ \sum_{n=2}^\infty\epsilon^n\int_0^{\epsilon^{-1/2}}f_n(Y)\,dY,
\label{eq:expand}
\end{eqnarray}
where $f_n$ are rational functions that increase in complexity with $n$. 
Integrating the above leads to 
 \begin{eqnarray}
\nonumber
\alpha+\pi&=\frac{1}{\epsilon}\left[\frac{2\sqrt{3+2\epsilon}}{1+\epsilon}\right]+\left[\frac{8}{3^{3/2}}\sinh^{-1}\left(\sqrt{\frac{3}{2\epsilon}}\right)-\frac{2+16\epsilon+8\epsilon^2}{3(\epsilon+1)^2\sqrt{2\epsilon+3}}   \right]\\ \nonumber&+\epsilon\left[\frac{10}{3^{5/2}}\sinh^{-1}\left(\sqrt{\frac{3}{2\epsilon}}\right)+\frac{18+41\epsilon-104\epsilon^2-189\epsilon^3-100\epsilon^4-20\epsilon^5}{9\epsilon(\epsilon+1)^3(2\epsilon+3)^{3/2}}\right]\\&+\dots
\label{eq:integrate}
\end{eqnarray}
where the 3 bracketed groups above correspond to the first 3 integrals of~(\ref{eq:expand}).  The full divergent behavior can be extracted from the first two bracketed groups as follows.  The first bracketed group in~(\ref{eq:integrate}) contains a rational function whose expansion about $\epsilon=0$ is a formal power series; thus this term leads to a $1/\epsilon$ divergence:
 \begin{eqnarray}
 \nonumber
\frac{1}{\epsilon}\left[\frac{2\sqrt{3+2\epsilon}}{1+\epsilon}\right]\sim\frac{2\sqrt{3}}{\epsilon}-\frac{4}{\sqrt{3}}+O(\epsilon),~\epsilon\to0.
\label{eq:term1}
\end{eqnarray}
The remaining bracketed groups shown (and those not shown) in~(\ref{eq:integrate}) each contain a $\sinh^{-1}$ term and a rational term.  Only the  $\sinh^{-1}$ terms, when expanded about $\epsilon=0$, ultimately lead to divergent behavior:    
 \begin{eqnarray}
\nonumber
\sinh^{-1}\left(\sqrt{\frac{3}{2\epsilon}}\right)\sim-\frac{1}{2}\ln\epsilon+\frac{1}{2}\ln6+O(\epsilon),~\epsilon\to0.
\label{eq:term2}
\end{eqnarray}
Only the second term of~(\ref{eq:integrate}) preserves this $\ln\epsilon$ divergence, as subsequent terms are multiplied by $\epsilon$ to some positive integer power.    The rational functions shown in~(\ref{eq:integrate}) evaluate to the following as $\epsilon\to0$
\begin{eqnarray}
\nonumber
\frac{2+16\epsilon+8\epsilon^2}{3(\epsilon+1)^2\sqrt{2\epsilon+3}}\sim\frac{2}{3^{3/2}}+O(\epsilon),~\epsilon\to0\\
\nonumber
\frac{18+41\epsilon-104\epsilon^2-189\epsilon^3-100\epsilon^4-20\epsilon^5}{9\epsilon(\epsilon+1)^3(2\epsilon+3)^{3/2}}\sim\frac{2}{3^{3/2}\epsilon}+O(\epsilon),~\epsilon\to0.
\end{eqnarray}
Note that both terms above, when substituted into~(\ref{eq:integrate}), lead to a constant due to a $\epsilon/\epsilon$ cancelation in the latter term. The remaining rational functions (not shown) in~(\ref{eq:integrate})  continue this pattern, each have a leading $1/\epsilon^n$ divergence as $\epsilon\to0$ that cancels with the $\epsilon^n$ pre-factors of the summation in~(\ref{eq:expand}), and thus will each evaluate to a constant + $O(\epsilon)$.  Substituting all asymptotic expressions above into~(\ref{eq:integrate}) leads to
 \begin{eqnarray}
\nonumber
\alpha+\pi&\sim\frac{2\sqrt{3}}{\epsilon}-\frac{4}{\sqrt{3}}+O(\epsilon)+\left[-\frac{4}{3^{3/2}}\ln\epsilon+\frac{4\ln6}{3^{3/2}}-\frac{2}{3^{3/2}}+O(\epsilon) \right]\\ &+\sum_{n=1}^\infty\epsilon^n\left[O(\ln\epsilon)+\frac{C_n+O(\epsilon)}{\epsilon^n} \right],~\epsilon\to0
\label{eq:term3}
\end{eqnarray}
where the following pattern emerges for the constants $C_n$:
\begin{equation}
C_{n+1}=\frac{n(2n+3)}{3(n+1)(n+2)}C_n,~C_1=\frac{2}{3^{3/2}}
\label{eq:constant}
\end{equation}
which may be written compactly as
\begin{equation}
C_{n}=\frac{1}{n}\left[\frac{2^{2-n}(2n+1)!}{3^{n+3/2}n!(n+1)!}\right].
\label{eq:constant}
\end{equation}
Equation~(\ref{eq:term3}) not only describes the dominant asymptotic behavior of $\alpha$ as $\epsilon\to0$, but also specifies the form of the next few subdominant higher-order terms $O(\epsilon\ln\epsilon)$, $O(\epsilon)$, $O(\epsilon^2\ln\epsilon)$; if the expansion is taken to higher order, this continues as $O(\epsilon^2)$, $O(\epsilon^3\ln\epsilon)$, $O(\epsilon^3)$, etc. Inserting~(\ref{eq:constant}) into~(\ref{eq:term3}) and truncating to the dominant asymptotic behavior leads to the following expression
\begin{eqnarray}
\nonumber
\alpha+\pi\sim&\frac{2\sqrt{3}}{\epsilon}-\frac{4}{\sqrt{3}}-\frac{4}{3^{3/2}}\ln\epsilon+\frac{4\ln6}{3^{3/2}}-\frac{2}{3^{3/2}}\\&+\sum_{n=1}^\infty\frac{1}{n}\left[\frac{2^{2-n}(2n+1)!}{3^{n+3/2}n!(n+1)!}\right]+O(\epsilon\ln\epsilon),~\epsilon\to0.
\label{eq:gather}
\end{eqnarray}
Note that the summation in~(\ref{eq:gather}) is convergent and can be written in closed form:
\[\sum_{n=1}^\infty\frac{1}{n}\left[\frac{2^{2-n}(2n+1)!}{3^{n+3/2}n!(n+1)!}\right]=\frac{8-4\sqrt{3}-8\ln\left[\left(3+\sqrt{3}\right)/6\right]}{3^{3/2}}.\]
Inserting the above into~(\ref{eq:gather}), using the substitution $\epsilon=2b'/(1+b')$, and simplifying like-terms leads to 
\begin{eqnarray}
\nonumber
\alpha+\pi\sim&\frac{\sqrt{3}}{b'}-\frac{4}{3^{3/2}}\ln b'+\frac{\sqrt{3}-4}{3}\\
&+\frac{4}{3^{3/2}}\ln\left(\frac{54}{6+3^{3/2}}\right)+O(b'\ln b'),~b'\to0.
\end{eqnarray}
As far as we are aware, this is the first time the full non-vanishing behavior of the bending angle in the strong-field limit has been established for a prograde orbit around an extremal ($a=1$) Kerr black hole.
 
\end{document}